\documentstyle[12pt]{article}
\def\p{pseudoscalar}
\def\a{\alpha}

\def\g{\gamma}

\def\e{\epsilon}

\def\ro{\rho}
\def\si{\sigma}
\def\A{{\rm\scriptscriptstyle A}}
\def\H{{\rm\scriptscriptstyle H}}
\def\T{{\rm\scriptscriptstyle T}}
\def\pt{p_\T}
\def\pa{p_\A}
\def\ga{g_\A}
\def\ma{M_\A}
\def\frac#1#2{{#1 \over #2}}
\def\slash#1{\rlap/#1}
\def\GeV{{\rm GeV}}
\def\TeV{{\rm TeV}}
\def\la{\langle}
\def\ra{\rangle}
\def\l{\left}
\def\r{\right}
\def\M{{\cal M}}
\def\o{\over}
\def\beq{\begin{equation}}
\def\beqa{\begin{eqnarray}}
\def\eeq{\end{equation}}
\def\eeqa{\end{eqnarray}}
\def\ch{\leftrightarrow}

\def\tr{{\rm tr}}
\def\no{\nonumber}
\def\noi{\noindent}
\def\qq{\qquad}
\def\re{{\rm Re}}
\def\etal{{\it et al.}}
\begin{document}
\pagestyle{empty}
\renewcommand{\thefootnote}{\fnsymbol{footnote}}
\begin{flushright}
MUHL-PHY/98-1 \\
 hep-ph/9808286
\end{flushright}
\vskip .5cm
\begin{center}
{\bf PRODUCTION OF A HIGGS PSEUDOSCALAR PLUS TWO JETS IN HADRONIC COLLISIONS}
 \\
\vskip .5cm
{\bf  Russel P.~Kauffman}\footnote{Email address: kauffman@muhlenberg.edu}
 \\
\vskip .5cm
{\it Department of Physics} \\
{\it Muhlenberg College, Allentown, PA~~18104} \\
\vskip .5cm
and \\
\vskip .5cm
{\bf  Satish V.~Desai}\footnote{Email address: sdesai@grad.physics.sunysb.edu}
 \\
\vskip .5cm
{\it Department of Physics} \\
{\it State University of New York at Stony Brook, Stony Brook, NY~~11794} \\
\vskip .5cm
ABSTRACT
\end{center}
\vskip 0.5cm
\noindent
We consider the production of a Higgs \p\ accompanied
by two jets in hadronic collisions.  
We work in the limit that the top quark is much heavier
than the Higgs \p\ and use an effective Lagrangian for the 
interactions of gluons with the \p.  We compute the amplitudes
involving: 1) four gluons and the \p, 2)
two quarks, two gluons and the \p\ and 3) 
four quarks and the \p.  
We find that the pseudoscalar amplitudes are nearly identical to those for
the scalar case,
the only differences being the overall size and the relative signs
between terms.  
We present numerical cross sections for proton-proton collisions with 
center-of-mass energy $14~\TeV$.  
\vskip0.25cm
\noindent August 1998\hfill\break
\noindent Submitted to {\it Physical Review D}
\vfill
\pagebreak
\pagestyle{plain}
\pagenumbering{arabic}
\section{Introduction}

The minimal model of electroweak symmetry breaking contains one complex
scalar doublet, three components of which become the longitudinal degrees
of freedom of the $W^{\pm}$ and $Z$.  The remaining component of the doublet
is the so-called Higgs boson.  However, the minimal model with one doublet
has no {\it a priori} justification and there are several motivations
for considering models with enlarged Higgs sectors, either containing 
more doublets, singlets or more exotic representations.  For example,
supersymmetric models require at least two doublets.  Likewise, at least
two doublets are required to produce CP violation in the Higgs sector
\cite{guide}.

Models with enlarged Higgs sectors have richer particle content than
the minimal model; in general, neutral pseudoscalars (with respect to
their fermion couplings)
and charged scalars
as well as extra neutral scalars are present. In this paper we will study
the production of a Higgs pseudoscalar ($A$) accompanied by two jets. 
Study of such processes gives information about the environment in which
the Higgs pseudoscalar is produced: additional jets may be used as tags or
be confused with decay products.  The production of additional jets also
influences the transverse momentum spectrum of the pseudoscalar.

Production of a Higgs pseudoscalar in hadronic collisions 
proceeds primarily via gluon fusion through a top-quark loop
(unless the coupling to bottom quarks is greatly enhanced). 
We will focus on the case of a light pseudoscalar and work in the 
heavy-top-quark limit : $m_t \gg \ma$.  In this limit, one can use
an effective Lagrangian to
couple the pseudoscalar to gluons.  The production of a Higgs pseudoscalar
plus one jet has been considered previously \cite{aonejet,spira}.
We compute all the contributions to the cross section for production of a pseudoscalar
plus two jets: $gg\to ggA$, $gg\to qqA$, $qg \to qgA$, 
$qq \to ggA$, $qq \to qqA$,
where `$q$' stands generically for a quark or anti-quark of undetermined
flavor. 

The organization of the paper is as follows.  The effective Lagrangian is
discussed in Section 2.  Section 3 contains the spinor product formalism
in which the amplitudes will be computed.  
The Higgs pseudoscalar plus four gluon amplitude is presented in 
Section 4.  Sections 5 and 6 contain the calculations of the amplitude 
involving a Higgs boson
plus a quark anti-quark pair and two gluons and the amplitude for a 
Higgs boson plus two quark anti-quark pairs, respectively.
Section 7 contains our numerical results and the Appendix
concerns the squares of the various amplitudes.

\section{The Effective Lagrangian}

We write the generic coupling of a quark to the Higgs pseudoscalar as
\break
$R_q m_q \g_5 /v$, where $R_q$ is a flavor/model dependent factor which we
will henceforth set equal to 1, $m_q$ is the mass of the quark, and 
$v=246~\GeV$ is the vacuum expectation value parameter.
The production mechanism in which we are interested is
$gg \rightarrow A$ which occurs
through a quark loop where the only contribution
which we will consider is that of the top
quark. In the limit in which
the top quark is heavy,
 $m_{\rm top} \gg \ma$, the cross section can be computed via the
following effective Lagrangian \cite{aonejet,spira}
\begin{equation} 
{\cal L}_{\rm eff} = \ga G_{\mu\nu}^a \tilde{G}_{\mu\nu}^a A,
\label{eq:leff}
\end{equation}
where $G^a_{\mu \nu}$ is the field strength of the SU(3) color
gluon field,
$\tilde{G}_{\mu\nu}^a$ is its dual, 
$\tilde{G}_{\mu\nu}^a = {1\o2}\e^{\mu\nu\rho\sigma} G_{\rho\sigma}^a$,
and $A$ is the pseudoscalar field.  The effective coupling 
is given by $\ga = \a_s /(8 \pi v)$.
The effective Lagrangian generates vertices involving the Higgs boson
and two or three gluons: 
\beqa
V_2(k_{1\mu}^a,k_{2\nu}^b) &=& -i \ga \delta^{ab} \e^{\mu\nu\rho\sigma}
                               k_{1\rho} k_{2\sigma},\no \\
V_3(k_{1\mu}^a,k_{2\nu}^b,k_{3\rho}^c) &=& -g\ga f^{abc} \e^{\mu\nu\rho\sigma}
                               (k_1 + k_2 + k_3)_\sigma, 
\label{eq:vertices}
\eeqa
where the $k_i$ are the gluon momenta directed outward and $a,b,c$ are their 
color indices.  The four-gluon vertex vanishes as it 
is proportional to the completely 
antisymmetric combination of structure constants:
\beqa
&&f^{abe}f^{cde} - f^{ace}f^{bde} + f^{ade}f^{bce} = \no \\
&&\qquad
-2 \tr \{ [T^a,T^b][T^c,T^d] - [T^a,T^c][T^b,T^d] + [T^a,T^d][T^b,T^c] \} = 0,\qquad
\eeqa
where the $T^i$ are the SU(3) generators.
It is straightforward to use this Lagrangian to obtain the ${\cal O}
(\alpha_s^3)$ contributions to the process $gg \rightarrow A$ 
\cite{aonejet,spira}.
These radiative corrections increase the 
lowest order rate by a factor of 1.5 to 2.  
Part of the calculation of the ${\cal O}(\alpha_s^3)$
 radiative
corrections to $gg\rightarrow A$ is the computation of  
the cross sections for $gg\rightarrow Ag $ and $gq \to Aq$ summed over spins.
One finds that 
these cross sections are identical to their counterparts for Higgs scalars, 
up to an overall factor.  Calculation of the helicity amplitudes for these
processes (using the spinor product formalism to be discussed below) yields 
the expected result that they too are equal to their counterparts, up to 
overall factors which include phases.

One might conjecture that the amplitudes for a Higgs pseudoscalar plus four
light partons would follow the same pattern.  However, the amplitudes 
involving the Higgs pseudoscalar must vanish in the limit that the momentum of
the Higgs pseudoscalar goes to zero.  
  This is readily seen from the structure 
of the three- and four-point vertices in Eq.~(\ref{eq:vertices}).  
They each can be rewritten to
be explicitly proportional to the momentum of the Higgs pseudoscalar.
The amplitudes for Higgs scalars can have
non-zero limits when the momentum of the Higgs scalar vanishes and so the pseudoscalar 
amplitudes must have a different form.

\section{Spinor Product Formalism}
We are interested in processes in which all the particles except
the Higgs \p\  are massless.  Each amplitude can be expressed in
terms of spinors in a Weyl basis.  For light-like momentum $p$ 
we introduce spinors \cite{helic,ber}
\begin{eqnarray}
&&|p{\pm}\ra ={1\over 2} (1\pm \gamma_5)u(p) = 
{1\over 2} (1\mp \gamma_5)v(p)\nonumber \\
&&\la p{\pm}| = \overline{u}(p){1\over 2} (1\mp \gamma_5)
= \overline{v}(p){1\over 2} (1\pm \gamma_5).
\label{eq:spindef}
\end{eqnarray}
Polarization vectors for massless vector bosons can be written in
terms of these spinors.  For a gluon of 
momentum $k$ and positive or negative helicity
\begin{equation}
  \epsilon^{\mu}_{\pm} = { \la q{\pm}|\gamma^{\mu}|k{\pm}\ra
                 \over \sqrt{2}\la q{\mp}| k{\pm} \ra } \quad,
\label{eq:epsilons}
\end{equation}
where the reference momentum $q$ satisfies $q^2=0$ and $q\cdot k\neq 0$
but is otherwise arbitrary.
Each helicity amplitude
can be expressed in terms of products of these spinors:
\begin{eqnarray}
\la p{-}|q{+} \ra &=& {-}\la q{-}|p{+} \ra \equiv \la pq \ra, \nonumber \\
\la p{+}|q{-} \ra &=& {-}\la q{+}|p{-} \ra \equiv [ pq ].
\label{eq:spinprod}
\end{eqnarray}
The identities
\begin{equation}
\slash p = |p{+}\ra \la p{+} | + |p{-}\ra \la p{-} | 
\label{eq:helproj}
\end{equation}
and
\begin{equation}
\la p {\pm} | \g^\mu | q {\pm} \ra \g_\mu
= 2 (|q{\pm}\ra \la p{\pm} | + |p{\mp}\ra \la q{\mp} | )
\label{eq:fierz}
\end{equation}
allow products of spinors and Dirac matrices to be written 
in terms of spinor products.

Amplitudes for processes involving the Higgs pseudoscalar contain expressions
of the form
$\e_{\mu\nu\rho\sigma}w^\mu x^\nu y^\ro z^\si$ where $w,x,y,z$ are
momenta, polarization vectors, or fermion currents.  These contractions can
be written in terms of spinor products through the following procedure.
We first write
\beq
\e_{\mu\nu\rho\sigma}w^\mu x^\nu y^\ro z^\si =
{1\o4i} \tr\{\slash w \slash x \slash y \slash z \g_5\}  
= {1\o4i} \tr\{\slash w \slash x \slash y \slash z (P_+ - P_-)\},
\label{eq:econtract}  
\eeq
where the projection operators are $P_\pm = (1\pm\g_5)/2$.
Using Eq.~(\ref{eq:helproj}) for momenta and Eq.~(\ref{eq:fierz}) for
polarization vectors and fermion currents,
each slashed vector can be written in terms of outer products of spinors:
\beq
\slash w = w_+|w_1{+} \ra \la w_2{+}| + w_-  |w_3{-} \ra \la w_4{-}|.
\label{eq:decompose}
\eeq
Inserting Eq.~(\ref{eq:decompose}) into Eq.~(\ref{eq:econtract})
and expressing the trace in terms of matrix multiplication, we have
\beq
\e_{\mu\nu\rho\sigma}w^\mu x^\nu y^\ro z^\si = {1\o4i} ( w_+ \la w_2{+}| \slash x \slash y \slash z | w_1{+} \ra
          - w_- \la w_4{-}| \slash x \slash y \slash z | w_3{-} \ra ),
\eeq
which reduces to spinor products upon application of Eq.~(\ref{eq:decompose})
to $\slash x$, $\slash y$, and $\slash z$.

For the remainder of the paper we will use the convention that all the 
particles are outgoing.  The amplitudes for the
various processes involving two incoming 
massless particles and two outgoing massless particles plus a Higgs \p\
can then be obtained by crossing symmetry. The momenta
of the massless particles are labeled $p_1,~p_2,~p_3,~p_4$ with the
Higgs \p\  momentum being $p_\A$.  Our convention is then
$p_1+p_2+p_3+p_4+\pa=0$. We will use the shorthand notations 
$\la p_i p_j \ra = \la ij \ra$, $ [p_i p_j ] = [ij]$, 
$(p_i + p_j)^2 = S_{ij}$, and $(p_i + p_j + p_k)^2 = S_{ijk}$.

\section{$Agggg$ Amplitude}

The $Agggg$ amplitude 
is obtained by summing the 25 
Feynman diagrams detailed in Fig.~1.   (The diagrams are the same as
the scalar case except for the absence of the vertex involving the \p\
and four gluons \cite{hgggg,hjj}.)
To facilitate the 
cancellations that simplify the amplitude we introduce the {\it dual
color decomposition}.
The scattering amplitude for a Higgs \p\ and $n$ gluons with external
momenta $p_1$, ...$p_n$, colors $a_1$,...$a_n$, and
helicities $\lambda_1$,...$\lambda_n$ is written as 
\cite{parki,parkii}
\beq
{\cal M}= 2 \ga g^{n-2}
\sum_{\rm perms} {\rm tr} (T^{a_1}...T^{a_n})m(p_1,\epsilon_1;
...;p_n,\epsilon_n),
\label{eq:dual}
\eeq
where the sum is over the non-cyclic permutations of the
momenta.
The ordered sub-amplitudes $m(p_1,\epsilon_1;...;p_n,\epsilon_n)$, which we 
abbreviate $m(1,...,n)$, are invariant under cyclic permutations of the
momenta, gauge transformations, and (modulo signs) reversal of the order of 
their arguments.  They also factorize in the soft and collinear limits.  
For $n=4$, the subamplitudes do not interfere in the sum over colors:
\beq 
\sum_{\rm colors} |\M|^2 = {g^2 \ga^2 \o 4} N^2 (N^2-1) 
\sum_{\rm perms} |m(1,2,3,4)|^2. 
\label{eq:incoh}
\eeq

The complete set of sub-amplitudes can be obtained from the following three:
\beqa
&&m(1^+,2^+,3^+,4^+)={\ma^4\over \la 1 2\ra \la 2 3\ra
   \la 3 4\ra \la 4 1\ra} 
\label{eq:pppp} \\ 
&&m(1^-,2^+,3^+,4^+) = -
{\la 1{-}|\slash \pa |3{-} \ra^2 [24]^2 \over S_{124} S_{12} S_{14}}     
-{\la 1{-}|\slash \pa | 4{-} \ra^2 [2 3]^2 \over S_{123} S_{12} S_{23}} \no\\
&&\qq
-{\la 1{-}|\slash \pa | 2{-} \ra^2 [3 4]^2 \over S_{134} S_{14} S_{34}} 
 +{[2 4] \over 
[ 1 2 ] \la 2 3\ra \la 3 4 \ra   [ 4 1]} 
\biggl\{ S_{23} {\la 1{-} |\slash \pa | 2{-} \ra \over \la 4 1\ra } \no \\
&&\qq\qq\qq\qq\qq\qq\qq +       S_{34} 
      {\la 1{-} |\slash \pa | 4{-} \ra \over \la 1 2\ra }
-[2 4 ] S_{234}\biggr\} 
\label{eq:mppp}\\ 
&&m(1^-,2^-,3^+,4^+)=-{\la 1 2\ra^4\over \la 12 \ra 
\la 23 \ra \la 34\ra \la 41\ra}
 + {[34]^4 \over [1 2] [23] [34] [41]} .
\label{eq:mmpp}
\eeqa
The structures containing $\pa$ can be expanded in terms of spinor products 
using Eq.~(\ref{eq:helproj}) and momentum conservation.  For example 
$\la 1{-}|\slash \pa |3{-} \ra = - ( \la 12 \ra [23] + \la 14 \ra [43] )$.
Permutations of $m(1^+,2^+,3^+,4^+)$ are obtained by permuting the 
momenta in the right side of Eq.~(\ref{eq:pppp}). Permutations of
$m(1^-,2^+,3^+,4^+)$ are obtained by permuting $p_2$, $p_3$ and $p_4$
in the right side of Eq.~(\ref{eq:mppp}) then using 
the cyclic and reversal properties
of the sub-amplitudes.  Permutations of
$m(1^-,2^-,3^+,4^+)$ are obtained by permuting the momenta in the 
{\it denominators} of the right side of Eq.~(\ref{eq:mmpp}) only. 
The amplitudes for the other helicity combinations can be
obtained (modulo phases) by parity transformations.  

Comparing these results to the scalar case \cite{hgggg,hjj}, we see a 
remarkable similarity. The results for the helicity combinations in which 
helicity is not conserved, ++++ and ${-}$+++, are identical to those for the 
scalar case and  the result for the helicity-conserving ${-}{-}$++ combination 
differs only in the relative sign between 
the two terms.  All three sub-amplitudes vanish in the limit that 
the momentum of the \p\ goes to zero.  The behavior of the ${-}$+++ amplitude 
in this limit is complicated by the fact that all the $S_{ijk}$'s vanish 
as well.  However, momentum 
conservation implies that, for example, 
$S_{123} = (p_4 + \pa)^2 \to 2 p_4\cdot\pa$ when $\pa\to0$.
limit.  Thus, the ${-}$+++ amplitude is proportional to one power of $\pa$.  
The ${-}{-}$++ amplitude vanishes because the two terms in Eq.~(\ref{eq:mmpp}) 
cancel in the limit $\pa\to0$ \cite{hjj}. 

\section{The $A  q \bar q g g$ Amplitude}
The $A  q \bar q g g$ amplitude 
can be obtained from the Feynman diagrams
of Fig.~2.  As was the case for the $Agggg$ amplitude, 
the calculation can be simplified by  judicious choice of color decomposition
\cite{kunszt,parkii}.
The amplitude for a Higgs \p, a quark--anti-quark pair with color indices
$i$, and $j$, and $n$ gluons with color indices $a_1,...,a_n$ can be 
written:
\beq
\M = -i g^n \ga
\sum_{\rm perms} (T^{a_1} T^{a_2} ... T^{a_n})_{ij} m(p1,\e_1;...;p_n,\e_n),
\label{eq:qdual}
\eeq
where the sum runs over all $n!$ permutations of the gluons and the 
sub-amplitudes $m(p1,\e_1;...;p_n,\e_n)$ have an implicit dependence on
the momenta and helicities of the quark and anti-quark.  For the case we
are interested in there are only two subamplitudes which we will label as
$m(3,4)$ and $m(4,3)$ since the gluon momenta are $p_3$ and $p_4$.
Like the subamplitudes for the pure gluon case these subamplitudes are
separately gauge independent and factorize in the soft gluon and collinear
particle limits.

Since they are on the same fermion line, the quark and anti-quark must have
opposite helicities.
Labeling the helicity amplitudes by the helicity of the quark, anti-quark
and the two gluons (in that order) we find
\beqa
m^{+-++}(3,4) &=& { \la 2{-}|\slash \pa| 3{-} \ra^2 \o S_{124} }
                 { [14] \o \la 24 \ra } \l({1\o S_{12}} + {1\o S_{14}} \r)
            -{ \la 2{-}|\slash \pa| 4{-} \ra^2 \o S_{123} S_{12}}
             { [13] \o \la 23 \ra }
\nonumber \\
            &+&{ \la 2{-}|\slash \pa| 1{-} \ra^2 \o [12] 
            \la 23 \ra \la 24 \ra \la 34 \ra }
\label{eq:qpmpp}
\eeqa
To get the subamplitude with the other ordering, $m^{+-++}(4,3)$, exchange
$p_3 \leftrightarrow p_4$ in this expression.  The other independent 
subamplitudes are
\beqa
m^{+-+-}(3,4) &=& {\la24\ra^3 \o \la12\ra \la23\ra \la34\ra } 
                 + { [13]^3 \o [12][14][34] }
\label{eq:qpmpmi} \\
m^{+-+-}(4,3) &=& -{ [13]^2 [23] \o [12][24][34] }
                  - {\la14\ra \la24\ra^2 \o \la12\ra \la13\ra \la34\ra } 
\label{eq:qpmpmii} 
\eeqa
The other helicity amplitudes (up to phases) can be obtained by parity, 
Bose symmetry,
and charge conjugation \cite{hjj}.  

Comparing to the results for the Higgs scalar we find the same pattern as in
the four-gluon case.  The helicity-violating case, +${-}$++, has the same 
subamplitudes as the scalar case, whereas the helicity conserving case,
+$-$+$-$, 
differs only by a relative sign between the terms.  Once again the 
subamplitudes vanish in the limit $\pa\to0$.    

\section{The $A q \bar q q\bar q $  Amplitude}

The remaining processes producing a Higgs \p\ plus two jets are those
involving a combination of four quarks and anti-quarks.  In the case where
the two pairs are of different flavors the amplitude can be obtained
from the Feynman diagram in Fig.~3.  In the case when the two pairs are
identical there is an additional diagram which can be obtained by switching
the $2 \ch 4$ in the diagram of Fig.~3.  We present the amplitude for the case
of two different quark pairs, since the identical case can be obtained from it.
The sole independent
helicity amplitude can be labeled in terms of the helicities of the 1st
quark, the 1st antiquark, the 2nd quark and the 2nd antiquark (in that order):
\beq
\M^{+-+-} = i \ga g^2 T^a_{ij}T^a_{kl} 
\l( {\la24\ra^2 \o \la12\ra \la34\ra} - { [13]^2 \o [12][34] } \r).
\label{eq:aqqqq}
\eeq
Note that this result is identical to that for the scalar case except for 
the relative sign between the two terms.  This sign difference causes the 
amplitude to vanish in the limit that the momentum of the \p\ vanishes.
The other helicity amplitudes can be obtained by parity and charge conjugation
transformations.

\section{Numerical Results and Conclusions}

We will present numerical results for the CERN Large Hadron Collider (LHC)
at a center-of-mass energy of $\sqrt{S}=14~\TeV$.
Since all the parton level cross sections 
are singular in the small $\pt$ limit of one of the 
jets we will place a $\pt$ cut on the outgoing jets.  Since there are also
collinear singularities we will require that the outgoing jets be separated 
by $\Delta R \equiv \sqrt{\Delta \phi^2+\Delta \eta^2} \ge 0.7$. 
We will also require the outgoing jets have rapidity $\mid y\mid <2.5$.
Since there are no singularities depending on the momentum of the Higgs \p\
we will allow it to be unconstrained, except for a $\pt$ cut.

The total cross section for production of a Higgs \p\ plus two jets is presented
in Fig.~4.  The dominant processes, which contribute roughly equally,
are $gg \to ggA$ and $q (\bar q) g\to q (\bar q) g$.  The cross section for each
process can be obtained approximately by rescaling the scalar result:
$\si_\A \simeq (A/\ga)^2 \si_\H = (9/64) \si_\H = (1/7.1) \si_\H$, where $A$ is the effective
scalar coupling $A = \a_s /8\pi v$ from Ref.~\cite{hjj}.  The deviations from this
approximate result range from a few per cent to 10\% for almost all the processes and 
for the total.  The sole exception is the process involving four quarks/antiquarks 
which deviates from the rescaling by 10\%-20\% but 
contributes negligibly to the total.  The
largest deviations occur at the largest values of the $\pt$ cutoff, as expected
since the rescaling is exact in the small-$\pt$ limit.

In summary, we have calculated the amplitudes for the production of a Higgs
pseudoscalar accompanied by two jets.  The calculation was performed in the heavy-top-quark limit
using an effective Lagrangian.  
The amplitudes for the helicity-violating processes are identical in form to those
for a scalar boson, differing only by an overall factor.  In addition to the overall
factor, the helicity-conserving amplitudes
have sign differences between terms which cause them
to vanish in the limit $\pa \to 0$.
We find that the cross section is around a few tenths of a
picobarn at the LHC.
Our results provide the four-dimensional part of the 
``real'' corrections to Higgs pseudoscalar production 
at non-zero transverse momentum.  They can be combined with the virtual corrections
to complete the next-to-leading order calculation.

\bigskip
\begin{center}
{\bf ACKNOWLEDGMENTS}
\end{center}
\medskip
The authors are grateful for the support of Franklin and Marshall College, 
where this project was begun.  R.K. thanks the Muhlenberg College Faculty 
Development and Scholarship Committee for financial support.  S.D. thanks the Research
Foundation of the State University of New York for financial support.

\section*{Appendix. Squaring the Amplitudes}

Since the pseudoscalar amplitudes are nearly identical in form to their scalar
counterparts, the results for the squares of the scalar amplitudes can be 
easily modified to apply to the pseudoscalar case.  Here we will only discuss
the differences between the scalar and pseudoscalar cases and refer the reader
to Ref.~\cite{hjj} for the scalar results.  

In the $Agggg$ process the only subamplitudes which differ from the scalar 
case are those with gluon helicities $--$++.  The two independent 
subamplitudes squared are:
\beqa
|m(1^-,2^-,3^+,4^+)|^2 &=& {S_{12}^3 \o S_{14} S_{23} S_{34}} 
			  +{S_{34}^3 \o S_{12} S_{14} S_{23}}
			  -{\{1234\}^2 - 2S_{12}S_{23}S_{34}S_{41}
			    \o S_{14}^2 S_{23}^2} \no \\
|m(1^-,3^+,2^-,4^+)|^2 &=& {S_{12}^4 + S_{34}^4 \o S_{13}S_{14}S_{23}S_{34}}
\no \\
&&- \Big[ (\{1234\}^2 {-} 2S_{12}S_{23}S_{34}S_{41})
     (\{1243\}^2 {-} 2S_{12}S_{24}S_{43}S_{31}) \no \\
&&\;\; +\{1234\}\{1243\}(\{1234\}\{1243\} {+}2S_{12}S_{34}\{1324\}) \Big] \no 
\\
&&  / [2(S_{13}S_{14}S_{23}S_{34})^2].  
\label{eq:g4mmppsq}
\eeqa
(Compare to Eq.~(A12) from 
Ref.~\cite{hjj}.) 

In the $Aq\bar qgg$ process the amplitude squared for helicities +$-$++ is 
given
by Eqs. (A13)-(A16) of Ref.~\cite{hjj} with the parameter $A$ replaced 
by $\ga$. 
The other independent helicity amplitude squared is given by 
Eq.~(A17) of Ref.~\cite{hjj} with the parameter $A$ replaced by $\ga$ and
with
\beqa
\lefteqn{
|m^{{+}{-}{+}{-}}(3,4)|^2 = {S_{13}^3\over S_{12} S_{14} S_{34}} 
+ {S_{24}^3\over S_{12} S_{23} S_{34}}
- {1\over S_{14} S_{23} S_{12}^2 S_{34}^2} }  \no\\
&&\qq\times\Big[-\{1243\}^2 \{1324\} - S_{13} S_{24} \{1234\} \{1243\} 
+ S_{12} S_{13} S_{24} S_{34} \{1324\}\Big] ,\no\\
\lefteqn{|m^{{+}{-}{+}{-}}(4,3)|^2 = 
{S_{13}^2 S_{23}\over S_{12} S_{24} S_{34}} 
+ {S_{14} S_{24}^2 \over S_{12} S_{13} S_{34}} - { \{1243\} \{1234\} 
+ S_{12} S_{34} \{1324\} \over S_{12}^2 S_{34}^2 } ,} \no\\
\lefteqn{2\re[m^{{+}{-}{+}{-}}(3,4)m^{{+}{-}{+}{-}}(4,3)^*]
= -{\{1324\}\over S_{12} S_{34}} 
\l({S_{13}^2\over S_{14} S_{24}}
+ {S_{24}^2\over S_{13} S_{23}} \r) } \no \\
&&\qquad\qquad\qquad\qquad\qquad\qquad - {2 \o S_{12}^2 S_{34}^2 } 
\big( \{1243\}^2 - 2 S_{12} S_{13} S_{24} S_{34} \big) .
\label{eq:qqgg2sub}
\eeqa
(When comparing to Eq.~(A18) of Ref.~\cite{hjj}, note that the third equation
is incorrect; it should contain a term opposite in sign to the second line
of the third equation in Eq.~(\ref{eq:qqgg2sub}) above.) 

For the $Aq\bar q q^\prime \bar q^\prime$ process, the square of 
Eq.~(\ref{eq:aqqqq}) gives
\beq
|\M^{+-+-}|^2 = {\ga^2 g^4 (N^2-1) \over 4 S_{12} S_{34}} 
\l[(S_{13} + S_{24})^2 - {\{1243\}^2\over S_{12} S_{34}}\r]. 
\label{eq:aqqqqsq}
\eeq
In the case of identical quark pairs, there is a second diagram whose square
can be obtained by switching $1\ch 3$ in Eq.~(\ref{eq:aqqqqsq}).  The 
interference term which arises is
\beq
-2\re[\M \M^*(1{\ch}3)] = {-A^2g^4 (N^2{-}1) \o 4N}
\l[{(S_{13} {+} S_{24})^2 \{1234\} 
{+} 2\{1324\} \{1243\}\over S_{12} S_{23} S_{14} S_{34}}\r].
\label{eq:aqqqqint}
\eeq
(Compare with Eqs. (A19) and (A20) from Ref.~\cite{hjj}.)

\section*{\bf Figure Captions}
\bigskip

\noi {\bf Fig.~1}.  The Feynman diagrams for the $ggggA$ amplitude.  Curly lines represent gluons
and the dashed lines represent the \p. 
There are
12 diagrams of type a), 3 of type b), 4 of type c), and 6 of
type d).
\medskip

\noi {\bf Fig.~2}.  The Feynman diagrams for the $q\bar q ggA$ amplitude.  There
is one diagram of type a), two of type b), 4 of type c) and one of type d).
\medskip

\noi {\bf Fig.~3}.  The Feynman diagram for the $q\bar q q^\prime \bar q^\prime A$
amplitude.  In the case when the quark pairs are identical there is a second
diagram with the quark lines switched.
\medskip

\noi {\bf Fig.~4}.  The cross section for production of a Higgs boson plus
two jets at the LHC for three values of the $\pt$ cut.
\medskip

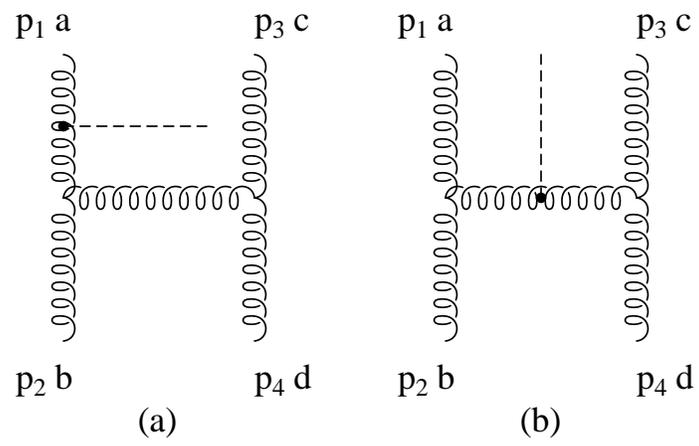
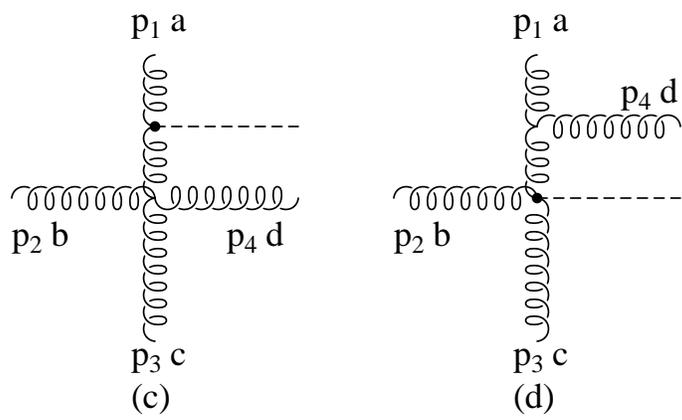
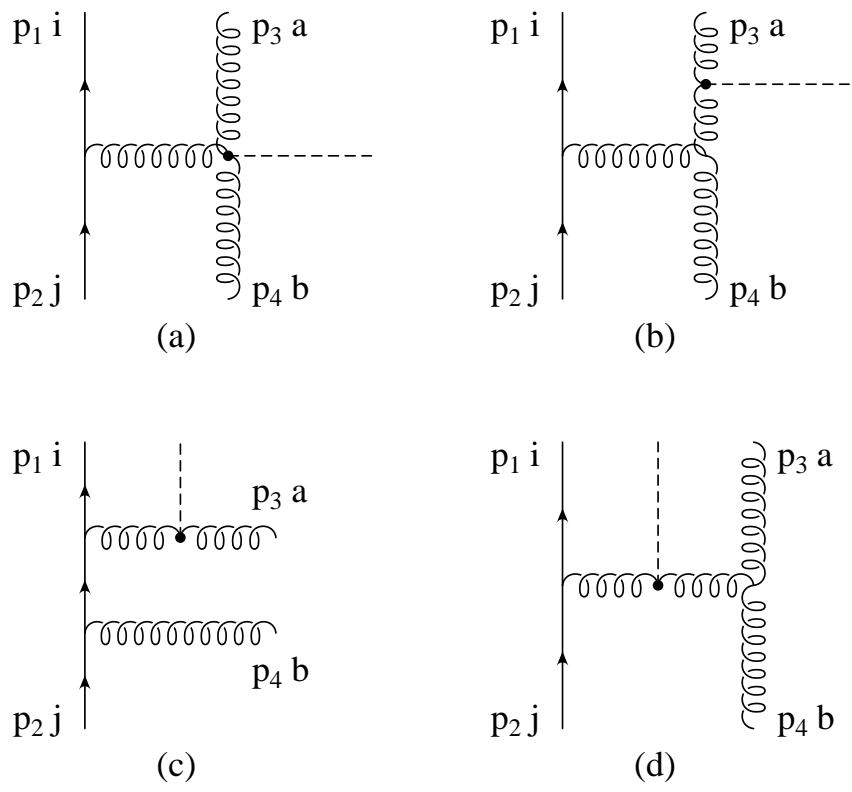
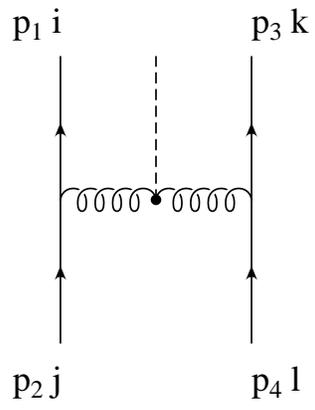
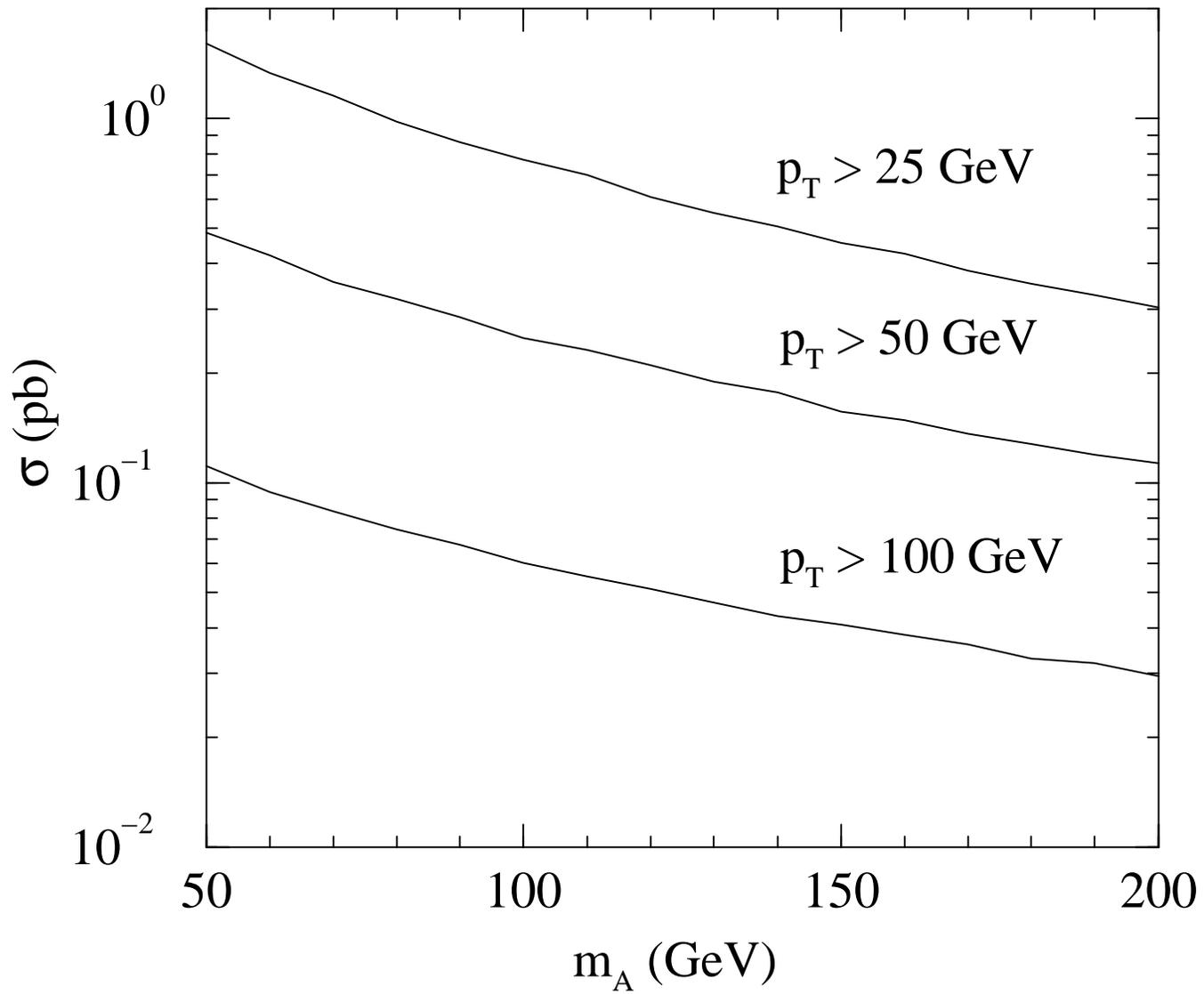
\end{document}